\documentstyle[12pt]{article}

\global\arraycolsep=1pt
\begin{document}

\renewcommand{\thefootnote}{\fnsymbol{footnote}}
\begin{titlepage}
%\null
\begin{flushright}
         NYU-THPH-9807\\
%   hep-th/9802180\\
January 31, 1998
\end{flushright}
\begin{center}
{\Large \bf Coulomb-gauge in QCD: \\renormalization and confinement}
\footnote{Lecture given at the 1997 Yukawa International Seminar (YKIS'97),
Yukawa Institute for Theoretical Physics, Kyoto, Japan, December 2-12, 1997}
\lineskip .75em
\vskip 3em
\normalsize
%{\large  Laurent Baulieu}\footnote{email address:
%baulieu@lpthe.jussieu.fr}   \\
%{\it LPTHE, Universit\'es Paris VI - Paris VII, Paris,
%France}\footnote{
%URA 280 CNRS,
%4 place Jussieu, F-75252 Paris Cedex 05, France.} \\
%and\\
%{\it SISSA, Via Beirut.2-4, 34013 Trieste, Italia}

%\vskip 1 em
{\large Daniel Zwanziger}\footnote{e-mail address:
daniel.zwanziger@nyu.edu}
\footnote{Work supported in part by National Science Foundation Grant no.
PHY9520978}\\  
{\it  Department of Physics, New York
University  \\  New York, NY 10003,
USA}
\end{center}
\vskip 1 em
\begin{abstract}
We review the Coulomb gauge in QCD and report some recent results. 
The minimal Coulomb gauge is defined and the fundamental
modular region, a region without Gribov copies, is described.  The Coulomb
gauge action is expressed in terms of the dynamical degrees of freedom with
an instantaneous Coulomb interaction, and its physical meaning is
discussed.  The local Coulomb gauge action with phase-space and ghost
variables is derived, and its BRS-invariance and renormalizability are
reviewed.  It is shown that the instantanteous part
$V(R)$ of $g^2 D_{00}(R, t)$, the time-time component of the gluon
propagator, is a renormalization-group invariant $V(R) =
f(R\Lambda_{QCD})/R$, and that the contribution of $V(R)$ to the Wilson
loop exponentiates.  It is conjectured that $V(R) \sim \kappa_{\rm coul}R$
at large $R$, and that 
$\kappa_{\rm coul}$ provides an order parameter for confinement of color
even in the presence of dynamical quarks.
\end{abstract}

\end{titlepage}
\renewcommand{\thefootnote}{\arabic{footnote}}
\setcounter{footnote}{0}

%%%%%%%%%%%%%%%%%%%%%%%%%%%%%%%%%%%%%%%%%%%%%%%%%%%%%%%%

\def\be{\begin{equation}}
\def\ee{\end{equation}}
\def\bea{\begin{eqnarray}}
\def\eea{\end{eqnarray}}
\def\F{gf}
\def\r{{\rm r}}
\def\rr{{\rm r}}

\def\I{ \Sigma_\r(\Phi,K_\Phi, V=0)}
\section{Introduction}
 
\newcommand{\ip}[2]{(#1, #2)}
                             % Defines \ip{arg1}{arg2} to mean
                             % (arg1, arg2).

%\newcommand{\ip}[2]{\langle #1 | #2\rangle}
                             % This is an alternative definition of
                             % \ip that is commented out.

 	Perturbatively, in each order, the Coulomb gauge must agree with
Lorentz-covariant gauges.  So why bother with the Coulomb gauge?

	Use of the Coulomb gauge is motivated by the conjecture \cite{Z1} that
non-perturbatively there is an instantaneous confining color-Coulomb
potential $V(r)$ which rises linearly $V(r) \sim \kappa_c r$ at large $r$. 
Here the color-Coulomb potential is defined by
\be
g_{(0)}^2 D_{00}^{(0)}(x) 
= g_{(0)}^2 \langle T A_0^{(0)}(x) A_0^{(0)}(0) \rangle
= V(r) \delta(t) + {\rm non \! - \! instantaneous},
\label{eq:Coulpot}
\ee
where $x^\mu = (t, \vec{x})$, $r = |\vec{x}|$, 
$A_\mu = (A_0,\vec{A})$, and the index $(0)$ refers to unrenormalized
quantities.  In the Coulomb gauge $\vec{A}$ is transverse,
$\vec{\nabla} \cdot \vec{A} = 0$. 
The quantity $\kappa_c$ is a string tension, characteristic of the Coulomb
gauge, that is finite in physical units.  

	Most of the results which will be reported here
are described in more detail in \cite{Z1}.  Among other things, it is shown
that
$V(R)$ is a renormalization-group invariant, and thus depends only on
physical masses, $V(r) = r^{-1} g_c^2(r\Lambda_{QCD})$.  Here 
$g_c(r\Lambda_{QCD})$ is a running coupling constant characteristic of the
Coulomb gauge, which is independent of the
renormalization mass $\mu$ and cutoff $\Lambda$, but which satisfies the
usual renormalization-group equation.

\section{Basics and Notation}

        The Yang-Mills field tensor is written
\be
F_{\mu\nu} = \partial_\mu A_\nu -   \partial_\nu A_\mu 
+ g_{(0)}A_\mu \times A_\nu,
\ee
where $(A_\mu \times A_\nu)^a \equiv f^{abc} A_\mu^b A_\nu^c$, and
$f^{abc}$ are the structure constants of the Lie algebra of SU(N).  The
classical field equations are given by 
\be
D^\mu(A) F_{\mu\nu} = j_{{\rm qu} \, \nu},
\ee
where
$D_\mu(A) \equiv \partial_\mu + g_{(0)}A_\mu \times$ is the gauge covariant
derivative in the adjoint representation, and the metric tensor has
signature $(-, +, +, +)$.  Under a local gauge transformation $g(x) \in$
SU(N), the connection 
$A_\mu = \lambda^a A_\mu^a$ transforms according to
\be
^g A_\mu = g A_\mu g^\dagger + g \partial_\mu g^\dagger.
\ee
Its infinitesimal form, for $g(x) = \exp\omega(x) \approx 1 + \omega(x)$,
is given by $^gA_\mu = A_\mu - D_\mu \omega$.

	The color-Gauss law
\be
\vec{D} \cdot \vec{E} = \rho_{\rm qu}    \label{eq:gauss}
\ee 
is a constraint that will be essential to the confinement picture in the
Coulomb gauge.

\section{Minimal Coulomb gauge}

	Gribov was the first to realize that the transversality condition 
 does not uniquely fix the gauge in
non-Abelian gauge theory \cite{Gribov}.  Thus, there exist
different transverse configurations $\vec{A}$ and
$\vec{A}_1$, with
$\vec{\nabla} \cdot \vec{A} = \vec{\nabla} \cdot \vec{A}_1 = 0$,
that are related by gauge transformation,
$\vec{A}_1 = \ ^g \vec{A}$.  Here
$\vec{A}$ and $\vec{A}_1$ are called ``Gribov copies'' of each other.  As
Gribov recognized, this has a significant effect on the dynamics because two
configurations that look very different may in fact be Gribov copies of each
other and thus represent the same physical configuration.  The physical
configuration space is the quotient space ${\cal P} = {\cal A} / {\cal G}$,
which is the space of configurations $\vec{A}$ modulo local gauge
transformations $g$.  It is gauge-invariant and is also known to be
topologically non-trivial (Singer's theorem) \cite{Singer}.  A complete
gauge fixing provides a concrete description of the
quotient space by means of representative configurations.  We now provide
such a description by means of the {\em minimal} Coulomb gauge.

	We shall fix a gauge by minimizing a conveniently chosen functional on
each ``gauge orbit'' \cite {Maskawa, Z2, Semionov}, a gauge orbit being the
set of all gauge transforms of one particular configuration $\vec{A}$. 
(``All'' must be suitably defined in the continuum theory.  However the
lattice analogs of quantities introduced below are
well defined, as is the minimal lattice Coulomb gauge
\cite{Z3, Z4}.)  The space of gauge orbits is the quotient space 
${\cal A} / {\cal G}$.

	A ``minimizing functional" for each time $t$ is introduced, which is the
restriction of the Hilbert norm to a generic gauge orbit,
\be
F_{\vec{A}}(g) \equiv \| \, ^g\vec{A} \, \|^2,
\ee
where
\be
\| \vec{A}\|^2 = \int d^3x \ |\vec{A}(t, \vec{x})|^2 .
\ee
is the Hilbert norm of $\vec{A}$ at time $t$.  To fix the
gauge, we choose as representative for each gauge orbit that configuration
for which $F_{\vec{A}}(g)$ achieves its absolute minimum.  The fundamental
modular region (FMR) is the set $\Lambda$ of representatives of all the
gauge orbits
\be
\Lambda \equiv \{A: F_{\vec{A}}(1) \leq F_{\vec{A}}(g) 
\; {\rm for \ all} \; g\}  \; .   \label{eq:FMR}
\ee
This is a region without Gribov copies.  Well, almost without Gribov
copies.  There are cases of ``accidental'' degeneracy, where the absolute
minimum is not unique.  The degenerate absolute minima are Gribov
copies that our gauge-fixing procedure has not eliminated.  One may show
that they occur only on the boundary of $\Lambda$.  These Gribov copies
must be identified topologically for $\Lambda$ to represent the quotient
space. 

	This specifies a complete gauge fixing, but it is not very explicit. 
However from the existence of the minimizing functional, one may deduce
explicit properties of the FMR.  Any minimum, which may be relative or
absolute, is a stationary point of  $F_{\vec{A}}(g)$, so
$\delta F_{\vec{A}}(1) / \delta g = 0$ for all $\vec{A} \in \Lambda$.  A
simple calculation gives 
$\delta F_{\vec{A}}(1) / \delta g = - 2 \vec{\nabla} \cdot \vec{A}$, and
we conclude that the Coulomb gauge condition 
$\vec{\nabla} \cdot \vec{A} = 0$ is satisfied for all $\vec{A} \in \Lambda$.
{\em In addition} there is a second condition which holds namely that  the
matrix of second derivatives of the
minimizing function is non-negative at an
absolute or relative minimum, and a simple calculation gives 
$ - \vec{D} \cdot \vec{\nabla} \geq 0$, where
$\vec{D} = \vec{D}(A)$.  Here the Faddeev-Popov operator
$M(A) \equiv - \vec{D} \cdot \vec{\nabla}$ has made its appearance.  It is
symmetric for $\vec{A}$ transverse, 
$ - \vec{D} \cdot \vec{\nabla} = - \vec{\nabla} \cdot \vec{D}$.  These two
conditions define the Gribov region $\Omega$.  It is the set of relative
and absolute minima of the minimizing function,
\be
\Omega \equiv \{\vec{A}: \ \vec{\nabla} \cdot \vec{A} = 0 \; {\rm and}  \;
- \vec{D} \cdot \vec{\nabla} \geq 0 \} ,
\ee
that contains the
fundamental modular region $\Lambda$ as a proper subset 
$\Lambda \subset \Omega$.  One may show that $\Lambda$ and $\Omega$ are
both convex, that both are bounded in every direction (this is proven
below), and that $\Lambda$ and $\Omega$ have some boundary points in common.

\section{Color-Gauss law}

	We have remarked that Gauss's law 
$\vec{D} \cdot \vec{E} = \rho_{\rm qu}$
is essential for the confinement  picture in the Coulomb gauge.  To solve
this constraint we pose 
$\vec{E} = \vec{E}^{\rm tr} - \vec{\nabla} \Phi$,
where $\Phi$ is the color-Coulomb potential, and 
$\vec{\nabla} \cdot \vec{E}^{\rm tr} = 0$.  Gauss's law reads
\be
- \vec{D} \cdot \vec{\nabla} \Phi = \rho.  \label{eq:solvegauss}
\ee
Here we have introduced a color charge density
\be
\rho \equiv - g_{(0)}A_i \times E_i^{\rm tr} + \rho_{\rm qu}  \label{eq:rho}
\ee
which includes only that part of the gluon charge density that comes
from~$\vec{E}^{\rm tr}$.  Gauss's law fixes $\Phi$,
\be
\Phi = ( - \vec{D} \cdot \vec{\nabla} )^{-1} \rho .
\ee
This equation, which appears neither gauge nor Lorentz covariant, in
fact holds in every gauge and every Lorentz frame.  However in the minimal
Coulomb gauge, $\vec{A}$ is transverse, $\vec{A} = \vec{A}^{\rm tr}$, so
$\rho$ depends only on the dynamical degrees of freedom of the gluon 
$A^{\rm tr}$ and $E^{\rm tr}$. 

	Moreover, as we have seen, the Faddeev-Popov
operator in the minimal Coulomb gauge, 
$M(\vec{A}) \equiv - \vec{D} \cdot
\vec{\nabla}$ is strictly positive in the
interior of the fundamental modular region, $M(\vec{A}) > 0$, so the
inverse which appears in
$\Phi = M^{-1}(A) \rho$ is well defined, except possibly for some points on
the boundary of $\Lambda$ where $M(\vec{A})$ may have a zero
eigenvalue.\footnote{Actually, $M(\vec{A})$ has a trivial null space
consisting of constant functions, 
$\vec{\nabla} \omega = 0$, and the inverse $M^{-1}$ is taken on the
orthogonal subspace.  Only $\vec{\nabla}\Phi$ appears in
$\vec{E} = \vec{E}^{\rm tr} - \vec{\nabla}\Phi$, so the arbitrariness of
$\Phi$ under
$\Phi \rightarrow \Phi + \omega$ is of no consequence.}  Thus
in the minimal Coulomb gauge there is a coherence between the gauge
fixing and the dynamics expressed by the Gauss law  constraint, in the
sense that this constraint may be solved by
inverting a positive operator.  

	The singularity of
$M^{-1}(\vec{A})$ strongly enhances the color-Coulomb potential
$\Phi$ as parts of the boundary of
$\Lambda$ are approached.  This is an expression of the famous
anti-screening property of QCD and, as we conjecture, of confinement. 
Indeed, the Green function which solves the color Gauss law constraint is
given by
\be
M^{-1}(\vec{A})
= ( - \vec{\nabla}^2 - g_{(0)}\vec{A}_i \times \vec{\nabla}_i )^{-1}
\ee
The first term $- \vec{\nabla}^2$ is strictly positive.  However the
second term has zero trace on the color indices, and thus has positive and
negative eigenvalues.  For some subspace there is a cancellation
between the first and second term, which gives a small denominator. 

	To see this in detail, consider the ray in \mbox{$A$-space} defined by
$\lambda \vec{A}$, where
$\lambda > 0$ is a positive parameter, and $\vec{A} \neq 0$ is arbitrary
but fixed \cite{Z5}.  For sufficiently small $\lambda$, the Faddeev-Popov
operator 
$M(\lambda \vec{A}) =  
- \vec{\nabla}^2 - \lambda g_{(0)}\vec{A}_i \times \vec{\nabla}_i$
is strictly positive.  As $\lambda$
increases, the lowest (non-trivial) eigenvalue of 
$M(\lambda \vec{A})$ decreases toward $0$, giving an enhanced color-Coulomb
potential, as we now show.  For let $\omega$ be a vector for which the
expectation value of the second term of $M(\vec{A})$ is negative,
$E \equiv (\omega, \ - g_{(0)}\vec{A}_i \times \vec{\nabla}_i \ \omega) <
0$.  Such a vector certainly exists because this operator has zero trace. 
We have
$(\omega, \ M(\lambda \vec{A}) \ \omega) = 
\| \vec{\nabla} \omega \|^2 + \lambda E$, with $E < 0$.  This expression
starts out positive and decreases linearly as $\lambda$ increases,
eventually becoming negative.  So for
$\lambda$ sufficiently large and positive, the lowest (non-trivial)
eigenvalue $\mu_0(\lambda \vec{A})$ of $M(\lambda \vec{A})$ is certainly
negative.  Consequently 
$\mu_0(\lambda \vec{A})$ is a continuous function of $\lambda$ that
starts out positive and goes negative, and thus it inevitably passes
through~0.  (One may show that the decrease is monotonic.)  When
$\mu_0(\lambda \vec{A}) = 0$, the point
$\lambda \vec{A}$ lies on the boundary $\partial\Omega$ of the Gribov
region~$\Omega$.  The color-Coulomb potential 
$\Phi = M^{-1}\rho$ is strongly enhanced as 
$\mu_0(\lambda \vec{A}) \rightarrow 0$. Incidentally, the above argument
also shows that the FMR is bounded in every direction~$\vec{A}$.

\section{Coulomb-gauge action in terms of dynamical degrees of freedom}

	A serious obstacle to the use of the Coulomb gauge in QCD has been doubt
about whether the theory is renormalizable and thus whether the gauge
really exists at all.  In this section we indicate the nature of the
problem and explain the solution physically by integrating out the
constraints so only the dynamical degrees of freedom remain.

	The Faddeev-Popov formula for the Euclidean functional
integral in the Coulomb gauge reads,
\be
Z = \int_\Lambda d^4A \ \delta(\vec{\nabla} \cdot \vec{A}) \  
\det ( - \vec{\nabla} \cdot \vec{D}) \ \exp( - S_{\rm cl}),
\label{eq:ZScl} \ee
where
\be
S_{\rm cl} = (1/2) \int d^4x \ (E^2 + B^2),
\ee
$E_i^a = F_{0i}^a$, and $B_i^a = F_{jk}^a$ for $i, j, k$ cyclic.  The
integration over $\vec{A}$ is restricted to the fundamental modular region,
$\Lambda$, but we temporarily ignore this point and discuss the
perturbative expansion which is insensitive to this restriction.

	The above expression certainly cannot be perturbatively renormalized in
the usual way, as one may see by considering the closed ghost loops that
represent the expansion of the Faddeev-Popov determinent.  The free
propagator of a ghost with momentum $\vec{k}$ is $1/\vec{k}^2$. 
It is independent of
$k_0$ and corresponds to the instantaneous space-time propagator 
$\delta(t)/r$, so a closed ghost loop is an instantaneous closed loop in
the Coulomb gauge. The insertion of the simplest ghost loop into a gluon
propagator of momentum $p$ gives the loop integral  
\be
\int dk_0 \int d^3k \ \frac{1}{\vec{k}^2 (\vec{p} - \vec{k})^2}.
\ee
The $k_0$ integration diverges because the integrand is independent of
$k_0$.  This occurs in any number of spatial dimensions, the so the
divergence cannot be regularized by dimensional regularization. 
Moreover the divergence is non-polynomial in $\vec{p}$.  One
suspects that this divergence is cancelled by other diagrams.  The problem
is to make this cancellation manifest.

	To do this we introduce the Gaussian identity
\be
1 = \int d^3\Pi \exp[ - \int d^4x \ (1/2) ( \vec{\Pi} + i\vec{E} )^2 ]
\label{eq:Gaussian} \ee
into $Z$, which gives
\be
Z = \int_\Lambda dA^{\rm tr} d^3\Pi dA_0 \  
\det ( - \vec{\nabla} \cdot \vec{D}) \ \exp( - S_1),
\ee
where
\be
S_1 \equiv \int d^4x \ [ i \vec{\Pi} \cdot \vec{E} + (1/2)(\Pi^2 + B^2)].
\ee
Here we have used the constraint expressed by 
$\delta(\vec{\nabla} \cdot \vec{A})$, to integrate out the
longitudinal part of $\vec{A}$, so $\vec{A} = \vec{A}^{\rm tr}$  in the
integrand, and the integation over 
$\vec{A}^{\rm tr}$ remains.
 
	With
$\vec{\Pi} \cdot \vec{E} = \vec{\Pi} \cdot 
(\partial\vec{A}^{\rm tr} / \partial t - \vec{D}A_0 )$, we observe that
the action is linear in $A_0$.  Integration over $A_0$ imposes the Gauss
law constraint,
$\delta(\vec{D} \cdot \vec{\Pi})$.  We separate the tranverse and
longitudinal parts of $\vec{\Pi}$ according to 
$\vec{\Pi} = \vec{\Pi}^{\rm tr} - \vec{\nabla} \Phi$, 
where $\Phi$ is the color-Coulomb potential, and we have
and
$d^3\Pi = {\rm const} d\vec{\Pi}^{\rm tr} \ d\Phi$.  
The Gauss law constraint reads 
$\delta( M \Phi - \rho)$,
where $M \equiv - \vec{D} \cdot \vec{\nabla}$ is the Faddeev-Popov
operator, and
\be
\rho \equiv - g_{(0)} A_i^{\rm tr} \times \Pi_i^{\rm tr}.
\label{eq:definerho} \ee  
We next effect the integration on $\Phi$ using
\be 
\int d\Phi \det M \ 
\delta( M \Phi - \rho) \ f(\Phi)
= f(M^{-1} \rho).
\ee
This absorbs the Faddeev-Popov determinent, and thereby achieves the goal of
eliminating the unwanted non-renormalizable instantaneous closed ghost
loops.  

	We obtain finally the partition function expressed as a functional integral
over the dynamical phase-space variables,
\be
Z = \int_\Lambda dA^{\rm tr} d\Pi^{\rm tr} \ \exp( - S_{\rm coul}),
\label{eq:ZScoul} \ee
where the Coulomb-gauge action is given by
\be
S_{\rm coul} \equiv \int d^4x \ 
\{ i \vec{\Pi}^{\rm tr} \cdot \partial\vec{A}^{\rm tr} / \partial t 
+ (1/2)[(\Pi^{\rm tr})^2 + (\vec\nabla M^{-1} \rho)^2 + B^2)] \}.
\ee
This formula is physically transparent.  The dynamical variables are the
canonically conjugate variables 
$\vec{A}^{\rm tr}$ and $\vec{\Pi}^{\rm tr}$ that remain after the
constraints are integrated out. This formula could also have been obtained
by the method of canonical quantization of constrained systems.  As
desired, there are no instantaneous closed loops because the Faddeev-Popov
determinent is gone.  There remains the instantaneous color-Coulomb
interaction
\be
S_{\rm inst} \equiv \int d^4x \ (\vec\nabla M^{-1} \rho)^2
= \int dt d^3y d^3z \  
\rho(t, \vec{y}) \ Q(\vec{y}, \vec{z}; \vec{A}^{\rm tr}(t)) 
\ \rho(t, \vec{z}),
\ee
with kernel  
\be
Q \equiv M^{-1} (-\vec\nabla)^2 M^{-1} .
\ee 

	This kernel contains the inverse
Faddeev-Popov operator $M^{-1}$ (twice). Recall that the lowest eigenvalue
of
$M$ approaches 0 as a generic point on the boundary $\partial\Omega$ of the
Gribov region~$\Omega$ is approached.  Consequently the Coulomb-gauge action
$S_{\rm coul}$ diverges as $\partial\Omega$ is approached, and
$\exp( - S_{\rm coul})$, regarded as a function of 
$g_{(0)}\vec{A}^{\rm tr}$, is
strongly suppressed by an essential singularity on $\partial\Omega$. 
The perturbative expansion in powers of $g_{(0)}\vec{A}^{\rm tr}$
reflects this strong suppression.  Therefore, although the perturbative
expansion is insensitive to the restriction to the fundamental modular
region $\Lambda \subset \Omega$, one may speculate whether
the dynamical suppression by $\exp( - S_{\rm coul})$ as $\partial\Omega$ is
approached, is sufficient for the perturbative expansion to capture in some
sense the non-perturbative features of the theory, for example by
resummation.

\section{Renormalizability of the local Coulomb-gauge action}

	Formula (\ref{eq:ZScoul}) is physically transparent, but it is unsuitable
for renormalization which requires a local action.  We therefore start over
and express the Faddeev-Popov formula (\ref{eq:ZScl}) in terms of the local
Faddeev-Popov action
\be
Z = \int d^4A dC d\bar{C} d\lambda   \exp( - S_{\rm FP}),
\label{eq:ZSFP} \ee
where
\be
S_{\rm FP} = \int d^4x [ \  (1/2)(E^2 + B^2) 
+ i\vec{\nabla} \bar{C} \cdot \vec{D}C 
+ i\vec{\nabla} \lambda \cdot \vec{A} \ ].
\ee
Here $C$ and $\bar{C}$ are the usual Faddeev-Popov ghosts and $\lambda$ is
a Lagrange multiplier that enforces the gauge constraint.  The instantaneous
closed fermi-ghost loops have been reintroduced, but we have seen that
cancellation of closed instantaneous loops is simpler in the phase space
representation.  We therefore insert the Gaussian identity eq.
(\ref{eq:Gaussian}) into the last expression and obtain
\be
Z = \int d^4A d^3\Pi dC d\bar{C} d\lambda   \exp( - S),
\label{eq:local} \ee
where
\be
S = \int d^4x [ \ i \vec{\Pi} \cdot \vec{E} + (1/2)(\Pi^2 + B^2) 
+ i\vec{\nabla} \bar{C} \cdot \vec{D}C 
+ i\vec{\nabla} \lambda \cdot \vec{A} \ ].  \label{eq:localaction}
\ee
We shall show that the instantaneous closed fermi ghost loops cancel
against the instantaneous closed bose loops. 

	Before doing so, we discuss some properties of this action.  The
equations of motion of $A_0$ and $\Pi$ read
\bea
\vec{D} \cdot \vec{\Pi} = 0  \nonumber \\
\vec{\Pi} = \vec{E} .
\eea
Thus in the local formulation of the Coulomb gauge,
Gauss's law is satisfied as an equation of motion, despite the  coupling to
ghosts.  This is the underlying reason why 
$g_{(0)}A_0^{(0)} = g_{(r)}A_0^{(r)}$ is invariant under
renormalization. 

	Consider the BRS transformation defined by
\bea
sA_\mu = - D_\mu C, \ \ \ \ \ \ \ sC & = & - (1/2)g_{(0)}C \times C
\nonumber \\
s\bar{C} = \lambda \ \ \ \ \ \ \ s\lambda & = & 0
\nonumber \\
s\Pi_i = C \times \Pi_i.
\eea
Here $A, C, \bar{C}, \lambda$ transform in the usual way, and $\Pi$
transforms covariantly, so $s$ is nilpotent, $s^2 = 0$.  Because $s$
effects an infinitesimal gauge transformation on $A_\mu$, the Yang-Mills
fields also transform covariantly,
$sE_i = C \times E_i$ and $sB_i = C \times B_i$.
The local action $S$ may be written
\be
S = \int d^4x [ \  (1/2)(\vec{\Pi} + i\vec{E})^2 
+ (1/2)(\vec{E}^2 + \vec{B}^2) 
+ s(i\vec{\nabla} \bar{C} \cdot \vec{A}) \ ],
\ee
which makes it obvious that $S$ is BRS invariant, $sS = 0$.  

	The only place in which a time derivative appears on the local action
(\ref{eq:localaction}) is in the term 
$i\vec{\Pi} \cdot \vec{E} 
= i\vec{\Pi} \cdot (\partial \vec{A}/\partial t - \vec{D}A_0)$, so from
Noether's theorem, the BRS charge density is 
$\rho_{\rm BRS} = -i \vec{\Pi} \cdot \vec{D}C$, and the conserved BRS
charge is simply
\be
Q_{\rm BRS} = i \int d^3x C \vec{D} \cdot \vec{\Pi}
= i \int d^3x C \delta S / \delta A_0,
\ee
which is intimately connected to Gauss's law.  Conservation of this charge
leads to a Ward identity which is specific to the Coulomb gauge,
\be
\int d^3x (\frac{\delta \Gamma}{\delta A_\mu} 
\frac{\delta \Gamma}{\delta K_\mu} + ...)
= \partial_0 \int d^3x ( C\frac{\delta \Gamma}{\delta A_0} + ... ).
\ee
Integration of this equation over all time annihilates the right hand side
and gives the familiar Zinn-Justin equation.  So this identity is more
detailed than the Zinn-Justin equation.  It expresses the constraint
imposed on renormalization by the color-Gauss law.  From it one may deduce
\be g_{(0)}A_0^{(0)} = g_{(r)}A_0^{(r)},
\ee
or in other words that $g_0A_0$ is a renormalization-group
invariant\cite{Z1}.  It follows that the time-time component of the gluon
propagator 
$g_{(0)}^2 D_{00}^{(0)}(R, t)$ is also a renormalization-group invariant,
as is its instantaneous part $V(R)$.  Thus it depends only on physical
masses, $V(R) = f(\Lambda_{QCD}R)/R$.

	The breaking of Lorentz invariance by the Coulomb gauge-fixing occurs only
in the BRS-exact piece of the action 
$s(i\vec{\nabla} \bar{C} \cdot \vec{A})$.  This allows one to control the
breaking of Lorentz invariance, and one may show that the
expectation value of a gauge-invariant quantity is Lorentz invariant.  

	There is no difficulty in deriving Feynman rules from the local
phase-space action $S$.  For example the $\Pi_i-\Pi_j$ propagator is 
$(\vec{k}^2 \delta_{ij} - \vec{k}_i\vec{k}_j)/k^2$,
the $\Pi_i-A_j$ propagator is
$k_0(\delta_{ij} - \vec{k}_i\vec{k}_j/\vec{k}^2 )/k^2$,
and the remaining propagators are as usual.

	We now show that the the instantaneous closed fermi loops cancel exactly
against instantanteous closed bose loops in every order of perturbation
theory.  The instantaneous closed fermi loops come from the term in the
action
$i\vec{\nabla} \bar{C} \cdot \vec{D}C$.  Now consider the term in the
action
\be 
i \vec{\Pi} \cdot \vec{E} = 
i(\vec{\Pi}^{\rm tr} - \vec{\nabla}\Phi) \cdot 
(\partial\vec{A}/\partial t - \vec{D}A_0),
\ee
where we have separated out the transverse and longitudinal parts of 
$\vec{\Pi}$.  The term 
$i\vec{\nabla}\Phi \cdot \vec{D}A_0$
has precisely the same form as the term as the ghost term
$i\vec{\nabla} \bar{C} \cdot \vec{D}C$.  For this reason it produces
instantaneous closed bose loops in every order of perturbation theory that
are equal, apart from sign, to the instantaneous closed fermi loops. 
Thus the non-renormalizable instantaneous closed loops cancel pairwise. 
Indeed this is why the ghosts are necessary.

	The last argument is not rigorous because it involves a cancellation of
infinite quantities.  However it may be made rigorous by use of the
interpolating gauge, defined by the gauge condition 
$- a\partial_0A_0 + \vec{\nabla} \cdot \vec{A} = 0$.  Here $a > 0$ is a
gauge parameter, and in the limit 
$a \rightarrow 0$, the interpolating gauge approaches the Coulomb gauge. 
For finite
$a$, the theory is renormalizable.  One may show \cite{BZ} that the
cancellation of the unwanted fermi and bose loops occurs at finite $a$,
modulo terms that vanish with $a$.  This establishes renormalizability of
the Coulomb gauge.

\section{Exponentiation of the static color-Coulomb potential}

	Consider a rectangular Wilson loop
\be
W = \langle {\rm tr \ P}\exp( \int g_{(0)}A_\mu^a \lambda^a dx^\mu ) \rangle
\ee
oriented along the time axis, of dimension $T \times R$.  The
anti-hermitian matrices $\lambda^a$ represent the SU(N) generators in some
representation D. \ In the Coulomb gauge, the time-time component of the
gluon propagator $g_{(0)}^2 D_{00}^{(0)}$ has an instantaneous piece
$V(r)\delta(t)$.  We shall evaluate the contribution $W_{\rm inst}$ to the
Wilson loop that comes from keeping only the instantaneous part of the gluon
propagator, without attempting to evaluate the remainder.  

	The instantaneous prt of the gluon propagator contributes all possible
ladder diagrams to the Wilson loop.  A typical ladder diagram, with time
ordering, 
\mbox{$0 , t_1 < t_2 < ...< T$}, contributes
\be
{\rm tr}(... \lambda^{a_2} \lambda^{a_1} \lambda^{a_1} \lambda^{a_2}...)
\int_{0 < t_1 < t_2 < ... < T} dt_1 dt_2 ... V(R) V(R)...  \ \ .
\ee
The path ordering for a ladder diagram produces the nested product
of $\lambda$-matrices that is shown.  The identity 
$\lambda^{a} \lambda^{a} = - C_D$, where $C_D$ is the Casimir for the
representation D, may be applied successively to the nested pairs of
contracted $\lambda$-matrices,
\be
{\rm tr}(... \lambda^{a_2} \lambda^{a_1} \lambda^{a_1} \lambda^{a_2}...)
= {\rm tr}(...) ( - C_D^2 ),
\ee
which allows us to evaluate all the traces.  The contribution
of instantaneous propagators to the Wilson loop exponentiates, giving
\be
W_{\rm inst} = (N^2 - 1) \exp[ - T C_D V(R)/2].
\ee

\section{Order parameter for color confinement}

	Recall that $V(R)$ depends only on
the physical mass $\Lambda_{QCD}$.  This result, and the exponentiation
formula for $W_{\rm inst}$, inspire the conjecture that $W_{\rm inst}$ obeys
an area law, or equivalently that $V(R)$ rises linearly at large $R$,
\be
V(R) \sim \kappa_{\rm coul} R.   \label{eq:kappacoul}
\ee
Here $\kappa_{\rm coul}$ is a string tension that is characteristic of the
Coulomb gauge, but finite in physical units.  

	It is believed, and confirmed by numerical simulation in lattice
gauge theory, that in gluodynamics the Wilson loop
in the fundamental representation obeys an area law 
$W \sim \exp( - \kappa TR )$, but that a perimeter law holds in the
adjoint representation.  Consequently, if eq. (\ref{eq:kappacoul}) is true,
it cannot possibly also be true that
$W \sim W_{\rm inst}$ in all representations D, for 
eq. (\ref{eq:kappacoul}) gives
an area law for $W_{\rm inst}$ whenever $C_D$ is non-zero.  
Nor does the Wilson loop $W$ obey an area law when
dynamical quarks are present, whereas we conjecture that in this case also
$V(R)$ rises linearly at large $R$.   This is as it should
be, because the color-Coulomb potential
$V(R)$ 
should not give the energy of a state which contains a pair of external
quarks, for that depends on whether mesons are formed.  Indeed the linear
rise of
$V(R)$ at large $R$ makes it energetically favorable in the presence of
an external quark for a dynamical anti-quark to be produced from the
vacuum, to form a meson, in which case the energy of a state containing an
external quark pair at large separation does not rise linearly with the
separation.  Rather the color-Coulomb potential
$V(R)$ couples universally to color-charge.  Thus if
$V(R)$ does rise linearly at large $R$, then
\be
\kappa_{\rm coul} = \lim_{R \rightarrow \infty} V(R)/R
\ee
serves as an order parameter for color-confinement, which is otherwise
lacking.  

	It appears that this conjecture is susceptible of investigation by
numerical simulation in lattice gauge theory.  Configurations that are
taken from the Wilson ensemble may be gauge fixed numerically to the
minimal Coulomb gauge, and $V(R)$ may then be found as the
expectation-value of instantaneous ($t = 0$) part of $g^2D_{00}(R, t)$.

\end{document}